\documentclass[aps, prd, twocolumn, showpacs, superscriptaddress, groupedaddress]{revtex4} 
\usepackage{graphicx}    
\usepackage{amssymb}
\usepackage{subfigure}
\usepackage{float}
\usepackage{ulem}
\usepackage[pagebackref=false, colorlinks=true]{hyperref}
\hypersetup{
linkcolor=blue,     
citecolor=blue,     
urlcolor=blue}   
\usepackage{soul}
\usepackage{amsmath}
\begin{document}
\title{On the visibility of singularities in general relativity and modified gravity theories}

\author{Karim Mosani}
\email{kmosani2014@gmail.com}
\affiliation{BITS Pilani K.K. Birla Goa Campus, Sancoale, Goa-403726, India}
\author{Dipanjan Dey}
\email{deydipanjan7@gmail.com}
\thanks{Corresponding author}
\affiliation{International Center for Cosmology, Charusat University, Anand 388421, Gujarat, India.}
\author{Pankaj S. Joshi}
\email{psjprovost@charusat.ac.in}
\affiliation{International Center for Cosmology, Charusat University, Anand 388421, Gujarat, India.}
\author{Gauranga C. Samanta}
\email{gauranga81@gmail.com}
\affiliation{P. G. Department of Mathematics, Fakir Mohan University, Balasore, Odisha, India.}
\author{Harikrishnan Menon}
\email{harikrishnanpmenon@gmail.com}
\affiliation{St. Xavier's College, University of Mumbai, Fort, Mumbai 400001, Maharashtra, India.}
\author{Vaishnavi D. Patel}
\email{vaishnavip1298@gmail.com}
\affiliation{PDPIAS, Charusat University, Anand 388421, Gujarat, India.}
\date{\today}

\begin{abstract}
We investigate the global causal structure of the end state of a spherically symmetric marginally bound Lemaitre-Tolman-Bondi (LTB) \cite{Lemaitre, Tolman, Bondi} collapsing cloud (which is well studied in general relativity) in the framework of modified gravity having the generalized Lagrangian $R+\alpha R^2$ in the action. Here $R$ is the Ricci scalar, and $\alpha \geq 0$ is a constant. By fixing the functional form of the metric components of the LTB spacetime, using up the available degree of freedom, we realize that the matching surface of the interior and the exterior metric are different for different values of $\alpha$. This change in the matching surface can alter the causal property of the first central singularity. We depict this by showing a numerical example. Additionally, for a globally naked singularity to have physical relevance, a congruence of null geodesics should escape from such singularity to be visible to an asymptotic observer for infinite time. For this to happen, the first central singularity should be a nodal point. We here give a heuristic method to show that this singularity is a nodal point by considering the above class of theory of gravity, of which general relativity is a particular case. 


\vspace{0.2cm}
\textbf{key words}: Gravitational Collapse, Naked Singularity, Higher-order gravity.

\end{abstract}
\maketitle

\section{Introduction}

A massive object contracting under the influence of its gravity can cause a singularity to form if its mass is beyond a certain limit 
\cite{Hawking}. 
This singularity, according to the weak cosmic censorship hypothesis (CCH) 
\cite{Penrose}, 
is hidden behind a null surface called the event horizon. However, lately, it has been shown that, in the case of a collapsing spherically symmetric dust cloud in the framework of General Relativity (GR),  sufficient radial inhomogeneity in the density can allow the null geodesics to escape the singularity and reach an observer far away, thereby making the singularity visible 
\cite{Deshingkar, Giambo, Jhingan, Mosani}. 
A singularity, if visible, can be classified into two types: local and global. In the case of locally visible singularity, the null geodesic can escape the singularity, but before crossing the boundary of the collapsing cloud, it comes across the trapped surfaces formed around the singularity and falls back in 
\cite{Joshi, Joshi2, JoshiCUP}. Since the observer must be inside the collapsing cloud to receive signals from only locally visible singularity, such a case may not have much astrophysical relevance. In the case of globally visible singularity, no such trapped surfaces come into the picture, and the outgoing singular null geodesics travels to future null infinity without any hindrance, and hence has more astrophysical relevance 
\cite{Mosani2}.

Coming to the theory of gravity, GR is a very successful theory and has been supported by various observational evidence like the measurement of the deflection angle of the light bent due to the curvature in the spacetime; which is predicted by GR as twice that predicted by Newton's theory (and correspondingly the gravitational time delay),
\cite{Watson}
the perihelion precession of the mercury, the gravitational redshift, etc
\cite{Schutz}. However, GR may not give a complete picture of the working of the universe. It has been found that GR is not renormalizable
\cite{Dewitt, Hooft, Goroff}.
It was, nevertheless, shown by Stelle 
\cite{Stelle} 
that actions with additional terms  $R^2$ and $R_{ij}R^{ij}$ in the Lagrangian can be renormalized. However, it had its shortcoming of unresolved unitarity problem. Additionally, as far as the $R_{ij}R^{ij}$ term is considered, Ostrogradski theorem does not allow its presence 
\cite{Woodard}.

$f(R)$ theories, however, do not contradict Ostrogradski’s result
\cite{Woodard}. 
In the strong gravity regime, higher-order terms of $R$ in action, if present, will dominate, and hence it may become essential to incorporate the modified action while investigating the neighborhood of the singularity formed due to gravitational collapse.
One such example of $f(R)$ gravity theory is the Starobinsky type gravity 
\cite{Starobinsky}, 
which was used to study curvature-driven inflationary scenarios. This model has the Lagrangian expressed as $R+\alpha R^2$, where $\alpha>0$ is a constant 
\cite{Starobinsky2}. 
The second-order curvature term $R^2$ can dominate and play a significant role in the strong gravity regime.

Because Starobinsky's theory is non-renormalizable, due to the absence of $R_{ij}R^{ij}$ term in the Lagrangian, every other higher-order term could also dominate near the singularities. For example, the effects of $R^3$ become even larger than that of $R^2$. One could thus argue that at such high-energy scales, one cannot simply truncate this infinite series and work at the level of $R^2$. \textcolor{black}{There are several papers in $f(R)$ and other modified gravity in which junction conditions have been discussed
\cite{Deruelle_2008, Senovilla_2013, Olmo_2021, Rosa_2021, Rosa_2022}). In 
\cite{Goswami}
it is shown that spatially homogeneous configurations cannot be matched smoothly with an exterior Schwarzschild spacetime in the domain of general form of $f(R)$ gravity.}
However, our concern in the present article is to see how the change in the Lagrangian affects the global causal property of the singularity. Additionally, this knowledge can be utilized to better understand the feature of a globally visible singularity in general. This is achieved in two folds : 

Firstly, using the $f(R)$ class of gravity, more specifically the Starobinsky type gravity, we address whether the first central singularity is a nodal point, which is a necessary property of the singularity to be physically relevant. This is done by identifying one outgoing singular null geodesic with each value of $\alpha$.

Secondly, dealing with the global causal structure of the singularity formed due to a collapsing cloud with non-zero internal pressure involves extreme complexity if one tries to approach the problem analytically in the framework of general relativity. This complexity is bypassed if one assumes zero pressure, which corresponds to LTB spacetime in general relativity. Whether the global causal structure of the singularity is generic in nature is a major concern, the knowledge of which is compromised if one only knows the outcome of the end state of the zero-pressure cloud collapse. One, therefore, has to know the end state of a non-zero pressured cloud collapse to have a say about the genericity of the end state of the gravitational collapse. In this direction of thought process, one could identify the existence of the globally visible singularity as an end state of LTB cloud in $R+\alpha R^2$ gravity with the existence of the globally visible singularity as an end state of a non-zero pressured collapsing cloud in $R+\alpha R^2$ gravity. It then seems judicious to conclude that global visibility of the singularity may not be unique to only vanishing pressure of the collapsing cloud in general, thereby seemingly making the causal structure of the singularity (global visibility) stable under a small perturbation in the initial data.

The article is organized as follows: In Section II, we give a brief overview of the Lemaitre-Tolman-Bondi (LTB) spacetime metric
\cite{Lemaitre, Tolman, Bondi} 
and the consequent collapse formalism, including the visibility property of the central singularity. This metric governs the inhomogeneous collapsing dust cloud in GR. In Section III, we discuss the $f(R)$ class of gravity theory
\cite{Odintsov, Sotiriou, Nojiri}, 
specifically the one with the Lagrangian having second-order curvature term. We ensure that the matter field governed by the LTB metric satisfies the strong, weak, null, and dominant energy conditions. The first central singularity is then investigated for its causal property. Comparisons are made between the visibility of the singularity formed due to two different matter fields, i.e., dust, and viscous fluid with heat flow, respectively in GR and in $f(R)$, both governed by the same LTB metric. However, the matching surface of the interior LTB metric and the exterior metric is different in different frameworks of gravity theory.  In the same section, we also discuss, using the $f(R)$ class of gravity, if the globally visible singularity is a nodal point. Finally, we end the article with the results and the conclusions drawn thereafter, in Section IV. Hereafter, we use the units wherein $c=8\pi G=1$.
\section{LTB geometry in general relativity and the collapse formalism}
In general relativity, the collapse of a spherically symmetric dust cloud is governed by the Laimetre-Tolman-Bondi (LTB) metric as follows:
\begin{equation}\label{LTB}
    ds^2=-dt^2+\frac{A'^2}{1+b}dr^2+A^2d\Omega^2, 
\end{equation}
in the comoving coordinate $t$ and $r$. Here, $A(t,r)$ is the physical radius of the collapsing cloud and is a monotone decreasing function of $t$, i.e., $\dot A<0$, $b(r)$ is called the velocity function and incorporates the information about the initial velocity of the collapsing cloud. The superscripts prime and dot denote the partial derivative with respect to radial and time coordinate, respectively.  The energy-momentum tensor for dust is expressed as
\begin{equation}
    T^{\mu \nu}=\rho U^{\mu} U^{\nu}.
\end{equation}
Here $U^{\mu}, U^{\nu}$ are the components of the four velocity. Using the Einsteins' field equation, one can get the expression of the density as
\begin{equation} \label{1efe}
\rho=\frac{S'}{A^2A'},
\end{equation}
where 
\begin{equation} \label{S}
  S=A\left(\dot A^2-b \right).
\end{equation}
$S$ is called the Misner-Sharp mass function
\cite{Misner}. 
$S/2$ gives the mass of the cloud inside the shell of radial coordinate $r$ at time $t$. Since the pressure inside the cloud is zero, the second Einsteins equation gives us
\begin{equation} \label{2efe}
    -\frac{\dot S}{A^2\dot A}=0,
\end{equation}
from which we can conclude that $S$ is independent of $t$, i.e the mass inside the shell of comoving radius $r$ is conserved. Due to the available degree of freedom, hereafter, we will consider the case for which $b=0$. This corresponds to a marginally bound collapse. The mathematical difficulties by taking such case are drastically reduced, although the case of non-marginally bound collapse ($b\neq 0$) can give results which are not very different qualitatively 
\cite{Mosani}. 
On integrating Eq.(\ref{S}), we obtain 
\begin{equation}\label{t-ts}
    t-t_s(r)=-\frac{2}{3}\frac{A^{\frac{3}{2}}}{\sqrt{S}}.
\end{equation}
Here $t_s(r)$ is the constant of integration, and is called the singularity curve. It tells us the time at which the shell of radial coordinate $r$ collapses to a singularity. Rescaling the physical radius using the coordinate freedom such that initially $A(0,r)=r$, one express the singularity curve as 
\begin{equation}\label{ts}
    t_s(r)=\frac{2}{3}\frac{r^{\frac{3}{2}}}{\sqrt{S}}.
\end{equation}
The causal structure of the singularity depends on the evolution of the apparent horizon. This horizon is the boundary of all the trapped surfaces. The expansion scalar for the outgoing null geodesic congruence, given by 
\cite{Poisson}
\begin{equation}\label{thetal}
    \Theta_l=h^{\mu \nu}\nabla_{\mu}l_{\nu}=\left (g^{\mu \nu}+\frac{l^{\mu}n^{\nu}+l^{\nu}n^{\mu}}{-l^{\alpha}n_{\alpha}}\right)\nabla_{\mu}l_{\nu}
\end{equation}
vanishes on this surface. On the trapped surface, this expansion scalar is negative.  $h^{\mu \nu}$ in the above equation is a spatial metric on the cross-section of the congruence, called the transverse metric; and $n^{\alpha}$ and $l^{\alpha}$ are the tangents of the incoming and outgoing null geodesics, respectively. \textcolor{black}{For a spherically symmetric space-time, one can say that on the apparent horizon $\Sigma$, at any point $p\in \Sigma$, the spin coefficient $\gamma_{\boldsymbol{10^{\prime}0}}^{\boldsymbol{~~~~1}}$ corresponding to a null frame must be zero and on the trapped surfaces the spin coefficient becomes negative. The unprimed and primed bold indexes of the spin coefficient are corresponding to the components of the spin bases (i.e. $i^{\boldsymbol{A}}, o^{\boldsymbol{A}}$) in the vector spaces and its conjugate vector spaces in the spin frame respectively.} It can be found from Eq.(\ref{thetal}) that on the apparent horizon, the Misner-Sharp mass function equals the physical radius of the collapsing cloud, i.e. 
\begin{equation}
   S(r)=A(t,r).  
\end{equation}
Using this, we can obtain the evolution of the apparent horizon, represented by the apparent horizon curve as
\begin{equation}\label{tah}
    t_{AH}(r)=\frac{2}{3}\frac{r^{\frac{3}{2}}}{\sqrt{S}}-\frac{2}{3}S
\end{equation}
for the marginally bound collapse. We can see that for $r=0$, the time of formation of the singularity and that of the apparent horizon is the same, keeping in mind the regularity of the initial data, i.e., $S(0)=0$. If the time of formation of the apparent horizon increases for increasing $r$, the outgoing singular null geodesics can escape. 

Depending on whether or not these geodesics later get trapped by trapped surfaces, one can obtain a locally or globally visible singularity, respectively. As mentioned in the Introduction, trapping of the outgoing singular null geodesic depends on the evolution of the event horizon, which is a null surface. The evolution of the event horizon is the same as the evolution of an infinitely redshifted null geodesic escaping the center (singular or non-singular center). Additionally, at the boundary of the collapsing cloud, the event horizon and the apparent horizon are indistinguishable. The dynamics of the EH is thus the solution of the null geodesic differential equation
\begin{equation}\label{deteh}
 \frac{dt_{EH}(r)}{dr}=A',   
\end{equation}
satisfying the condition 
\begin{equation}\label{teh}
t_{EH}(r_c)=\frac{2}{3}\frac{{r_c}^{\frac{3}{2}}}{\sqrt{S(r_c)}}-\frac{2}{3}S(r_c),    
\end{equation}
which is obtained from Eq.(\ref{tah}). The necessary condition for the singularity to be globally visible is that at $r=0$, $t_{EH}=t_s$. 

If the event horizon at the center forms before the formation of the singularity, then, even if there exists an outgoing singular null geodesic with a positive tangent at $r=0$, it will later get trapped by the trapped surfaces and fall back to the singularity, thereby making the singularity visible only locally. It should be noted that the event horizon at $r=0$ can not form after the formation of the singularity at $r=0$. This is because no null geodesic wavefront starting from $r=0$ can evolve further if formed after $t_s(0)$ because of the existence of trapped surface already present surrounding the center, as seen from Eq.(\ref{tah}). It has been shown before that there exists a non-zero measured set of initial data, $S$ and $b$, in the LTB case, leading to a globally visible singularity. 

In the Starobinsky type gravity, i.e., the one where we use the generalized lagrangian $f(R)= R+\alpha R^2$ in the action, we can show that the LTB metric is a solution, i.e., the matter field governed by such metric satisfies all the energy conditions. Unlike in GR, the LTB metric does not hold any special importance in this modified theory of gravity. However, for the sake of proving that the causal structure of the singularity is not a purely geometric property, we consider the same LTB metric in the $f(R)$ gravity to highlight the difference in the visibility of the singularity formed in this case from that in GR, in the next section.


\section{Gravitational collapse in f(R) gravity}
The total action for $f(R)$ gravity is written as
\cite{Sotiriou}
\begin{equation}
    S=\frac{1}{2}\int f(R) \sqrt{-g}d^4x+ S_m,
\end{equation}
where $S_m$ is the matter lagrangian. Using variational principle with respect to the metric gives the following field equation:
\begin{equation}
    G_{\mu \nu}=\frac{1}{F(R)}\left(T^{(m)}_{\mu \nu}+T^{(D)}_{\mu \nu}\right),
\end{equation}
where $F(R)=\frac{df(R)}{dR}$, and 
\begin{equation}
T^{(D)}_{\mu \nu}=\frac{(f-RF)}{2}g_{\mu \nu}+\nabla_{\mu}\nabla_{\nu}F-g_{\mu \nu}\Box F    
\end{equation}.
Here $\Box=g^{i j}\nabla_i \nabla_j$. In $f(R)$ theories of gravity, from the above field equation, even in the absence of a matter field, the Einstein tensor can be non-zero, unlike in GR. One can interpret that $T^{(D)}_{\mu \nu}$  be considered as the energy-momentum tensor, which has a purely geometric origin.  However, it is not under obligation to obey the strong energy conditions.

Due to the existence of a pure geometric part ($T^{(D)}_{\mu \nu}$) in the effective energy-momentum tensor in $f(R)$ gravity, the focusing properties of null-geodesic congruence differs from that in Einstein's general relativity. As we know, to investigate the focusing properties of null geodesic congruence in space-time, we need to use the \textcolor{black}{Raychaudhuri} equation for null geodesics which can be written as \cite{Poisson},
\begin{equation}
\frac{d\Theta_l}{d\lambda}=-\frac12 \Theta_l^2 - \sigma^{\alpha\beta}\sigma_{\alpha\beta}+\omega^{\alpha\beta}\omega_{\alpha\beta}-R_{\alpha\beta}l^{\alpha}l^{\beta}\,\, ,
\label{Ray1}
\end{equation}
where $\sigma_{\alpha\beta}$ is the shear tensor and $\omega_{\alpha\beta}$ is the rotation tensor and they are orthogonal to the null vector $l^{\alpha}$. For hyper-surface orthogonal (i.e. $\omega_{\alpha\beta} = 0$) null geodesic congruence, an initially converging ($\Theta_l(\lambda=\lambda_0)=\Theta_0<0$) null geodesic congruence forms a caustic point (where $\Theta_l\to -\infty$) within an affine parameter $\lambda\leq \frac{2}{|\Theta_0|}$, if $R_{\alpha\beta}l^{\alpha}l^{\beta}>0$ i.e. the null energy condition holds. In Einstein's general relativity, $R_{\alpha\beta}l^{\alpha}l^{\beta}>0$ implies $T^{(m)}_{\alpha\beta}l^{\alpha}l^{\beta}>0$; However, in $f(R)$ gravity, 
\begin{equation}
R_{\alpha\beta}l^{\alpha}l^{\beta}=T^{(m)}_{\alpha\beta}l^{\alpha}l^{\beta}+l^{\alpha}l^{\beta}\nabla_{\alpha}\nabla_{\beta}F\, ,
\label{nullcon}
\end{equation}
which implies that for a hyper-surface orthogonal null geodesic congruence, $\frac{d\Theta_l}{d\lambda}<0$ when $T^{(m)}_{\alpha\beta}l^{\alpha}l^{\beta}+l^{\alpha}l^{\beta}\nabla_{\alpha}\nabla_{\beta}F> 0$. This difference affects the dynamics of the apparent horizon and trapped surfaces inside a collapsing body and therefore, it may change the causal structure of the singularity formed in a gravitational collapse. If we consider LTB space-time as the solution to the $f(R)$ gravity, then the extra term in the above equation can be written as:
\begin{eqnarray}
l^{\alpha}l^{\beta}\nabla_{\alpha}\nabla_{\beta}F = \alpha\left(\partial_0^2R+2\frac{\sqrt{1+b(r)}}{A^{\prime}}\mathcal{U}+\frac{1+b(r)}{A^{\prime 2}}\mathcal{W}\right)\,\, ,\nonumber\\
\end{eqnarray} 
where $\mathcal{U}=\left(\frac{\dot{A}^{\prime}R^{\prime}}{A^{\prime}} + \dot{R}^{\prime}\right)$,\\ $\mathcal{W}=\left(R^{\prime\prime}+\frac{\dot{A}^{\prime}A^{\prime}\dot{R}}{1+b(r)}+\frac{2A^{\prime\prime}+2b(r)A^{\prime\prime}-b^{\prime}A^{\prime}}{2(1+b(r))A^{\prime}}R^{\prime}\right)$ and as mentioned before we consider $f(R)= R+\alpha R^2$. The Ricci scalar ($R$) is given in terms of the metric components and their derivatives as
\begin{equation}
    R=2\left(\frac{\ddot A'}{A'}+\frac{\dot A^2}{A^2}+\frac{2\dot A \dot A'}{A A'}+\frac{2\ddot A}{A}\right).
\end{equation}
In the above expression of $l^{\alpha}l^{\beta}\nabla_{\alpha}\nabla_{\beta}F$, the components of the null vector $l^{\mu}$ in the coordinate basis are
\begin{equation}
l^{\mu}=\{\frac{1}{\sqrt{2}},\frac{\sqrt{1+b(r)}}{\sqrt{2}A^{\prime}},0,0\}\,\, .
\end{equation}
After analyzing all the terms of the Eq.~(\ref{nullcon}), one can understand how the dynamics of the congruence of null geodesics in LTB space-time in the realm of $f(R)$ gravity differs from that in Einstein's general relativity. 
\textcolor{black}{Using the above form of null vector $l^\mu$ and the LTB metric in Eq.~(\ref{LTB}), in the realm of Einstein's general relativity, we can write down the Raychaudhuri equation (Eq.~(\ref{Ray1})) as,
\begin{eqnarray}
    \frac{d\Theta_l}{d\lambda}=l^\mu\nabla_\mu\Theta_l=-\frac12 \Theta_l^2+\kappa \Theta_l-\frac{S'}{2A^2A'} \,\, ,
\end{eqnarray}
where non-affinity coefficient $\kappa=n^\mu l^\nu \nabla_\nu l_\mu$, $R_{\mu\nu}l^\mu l^\nu = \frac{\rho}{2}=\frac{S'}{2A^2A'}$, and $n^\mu$ denote the components of another linearly independent null vector in the coordinate basis, and having the following relation with $l^\mu$: $l^\mu n_\mu = -1$. On the other hand, in $f(R)$ gravity the above equation becomes:
\begin{eqnarray}
    \frac{d\Theta_l}{d\lambda}=-\frac12 \Theta_l^2+\kappa \Theta_l-\frac{S'}{2A^2A'}+l^{\alpha}l^{\beta}\nabla_{\alpha}\nabla_{\beta}F\,\, .
\end{eqnarray}
The signature of the last term in the above equation plays an important role in the congruence of null geodesics in LTB spacetime in the realm of $f(R)$ gravity. The extra term is a manifestation of the $T^{(D)}_{\mu \nu}$ part of the effective energy-momentum tensor due to the geometric property of $f(R)$ gravity. Therefore, that extra part of the energy-momentum tensor in $f(R)$ gravity changes the dynamics of the congruence of null geodesics that can be seen in Einstein's general relativity.}
\subsection{Collapsing matter field and the energy conditions}
For the theory of gravity where the generalized Lagrangian in the action is $f(R)=R+\alpha R^2$, the  collapsing cloud governed by the  LTB metric, as shown in Eq.(\ref{LTB}) is a viscous fluid with heat flow, unlike in GR, as seen in the equation:
\begin{equation}
    T^{\mu \nu}= \rho U^{\mu}U^{\nu}+ph^{\mu \nu}+2q^{(\mu}n^{\mu)}-\Pi \left(n^{\mu}n^{\nu}-\frac{1}{3}h^{\mu \nu}\right).
\end{equation}
Here,
\begin{itemize}
    \item $h_{\mu \nu}$ is the transverse metric, expressed as
\begin{equation}
    h_{\mu \nu}=U_{\mu}U_{\nu}+g_{\mu \nu},
\end{equation}

\item $n^{\mu}$ is a spatial unit vector in the radial direction satisfying 
\begin{equation}
n_{\mu}n^{\mu}=1 \hspace{0.2cm} \textrm{and} \hspace{0.2cm} n_{\mu}U^{\mu}=0.
\end{equation}
\item $q^{\mu}$ is the heat flux vector and is spacelike, i.e. 
\begin{equation}
    q_{\mu}U^{\mu}=0.
\end{equation}
It describes the heat conduction such that $q_{\mu}n^{\mu}$ is the heat, crossing a unit surface which is perpendicular to $n^{\mu}$, per unit time. 
\item $p$ is the effective pressure given by
\begin{equation}
    p=\frac{p_r+2p_t}{3},
\end{equation}
where $p_r$ and $p_t$ are the radial and tangential components of the pressure inside the collapsing cloud.
\item $\Pi$ measures the anisotropy in the pressure given by
\begin{equation}
    \Pi=p_t-p_r.
\end{equation}
\end{itemize}
\begin{figure*}\label{fig2}
\subfigure[]
{\includegraphics[scale=0.68]{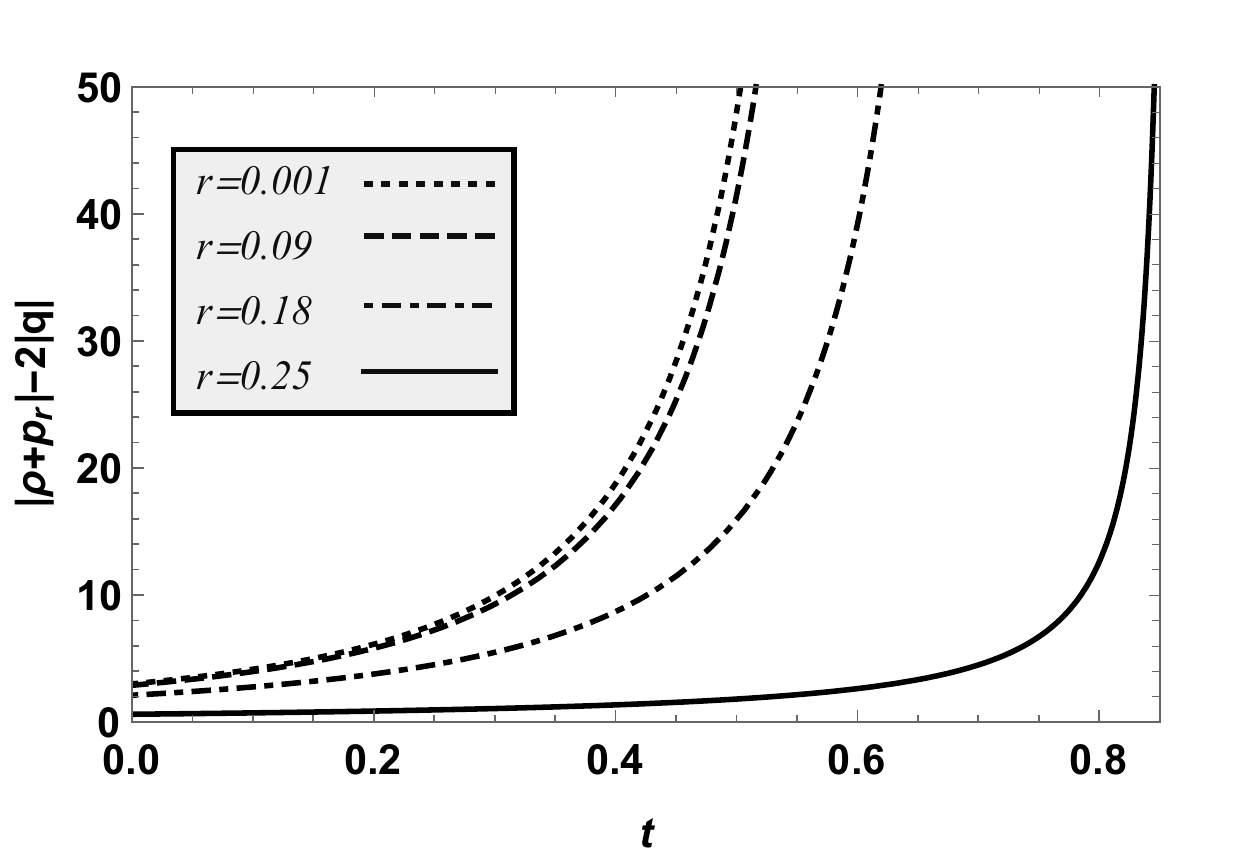}}
\hspace{0.2cm}
\subfigure[]
{\includegraphics[scale=0.68]{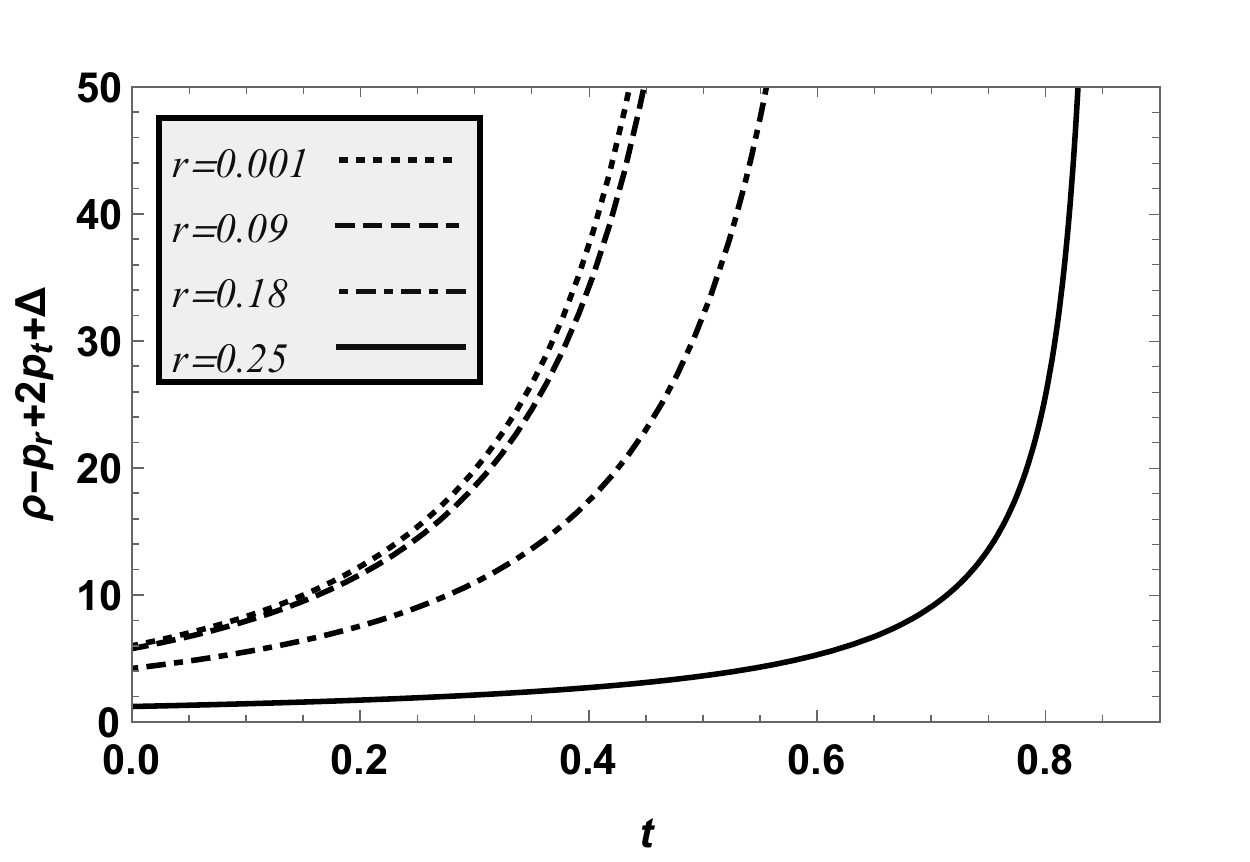}}
\hspace{0.2cm}
\subfigure[]
{\includegraphics[scale=0.68]{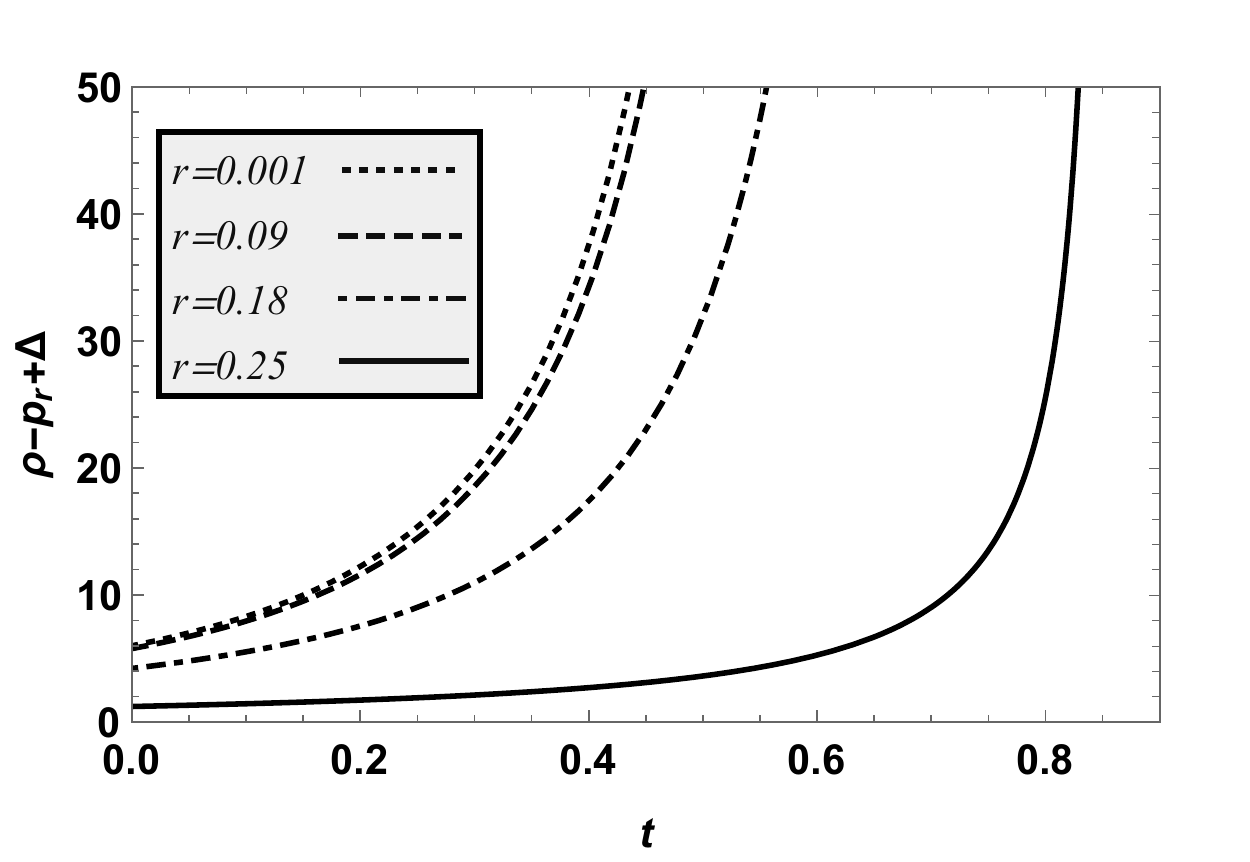}}
\subfigure[]
{\includegraphics[scale=0.68]{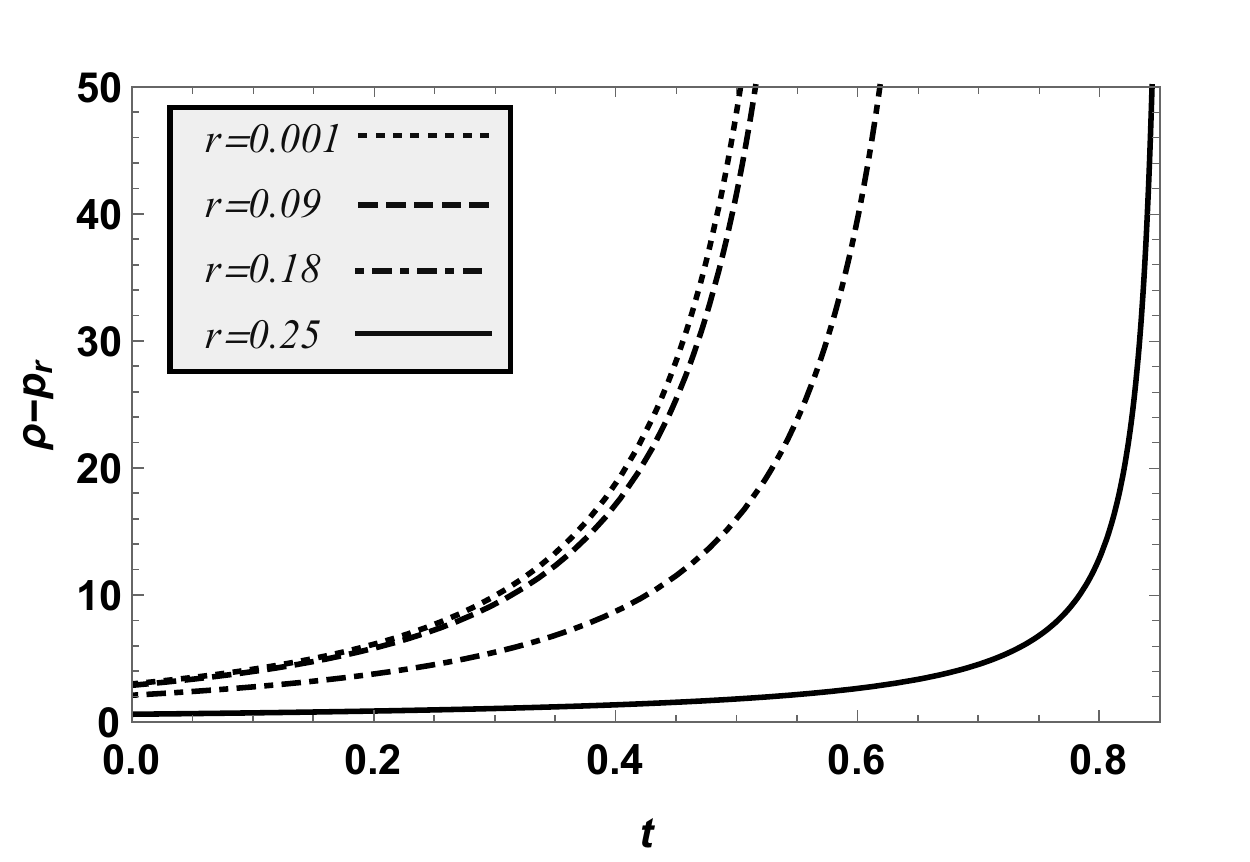}}
\hspace{0.2cm}
\subfigure[]
{\includegraphics[scale=0.68]{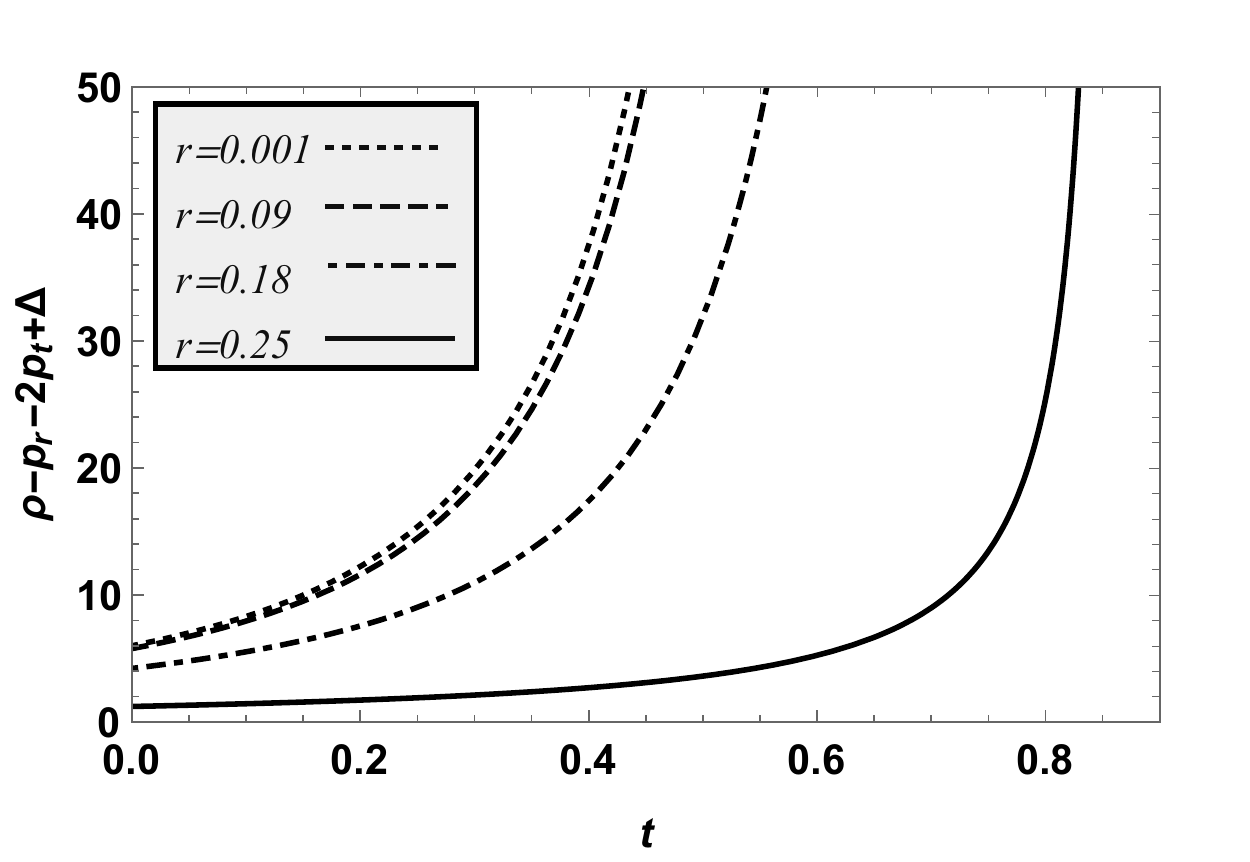}}
\hspace{0.2cm}
\subfigure[]
{\includegraphics[scale=0.68]{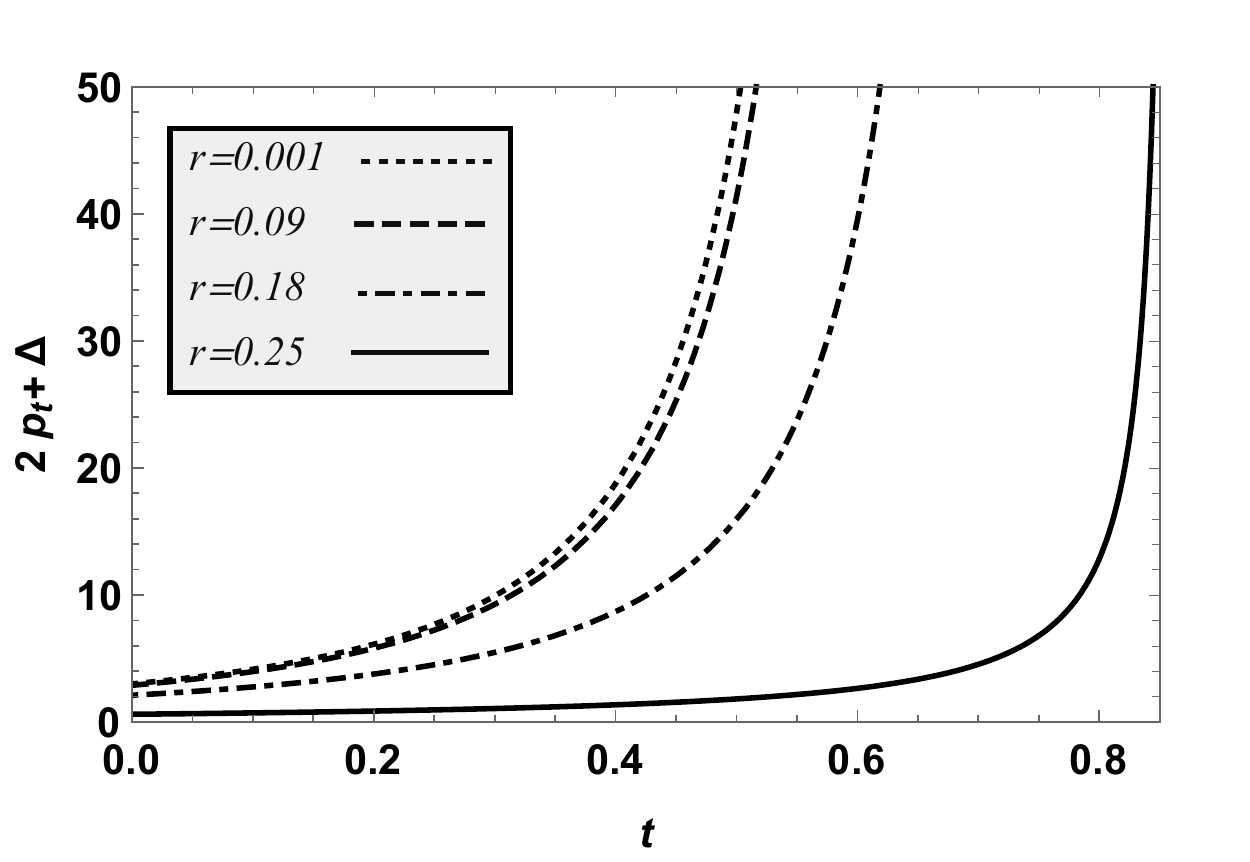}}
\hspace{0.2cm}
\caption{Various energy conditions of the collapsing matter cloud for different comoving radius $r$, throughout the collapse, is depicted here in the framework of $f(R)=R+\alpha R^2$ gravity with $\alpha=10^{-6}$. The initial data is taken as $S=r^3-25.5 r^6$. It can be seen from here that all the energy conditions (the inequalities in Eq.(\ref{ec1}-\ref{ec6})) are satisfied.}
\end{figure*}

\begin{figure*}\label{figure2}
\subfigure[]
{\includegraphics[scale=0.54]{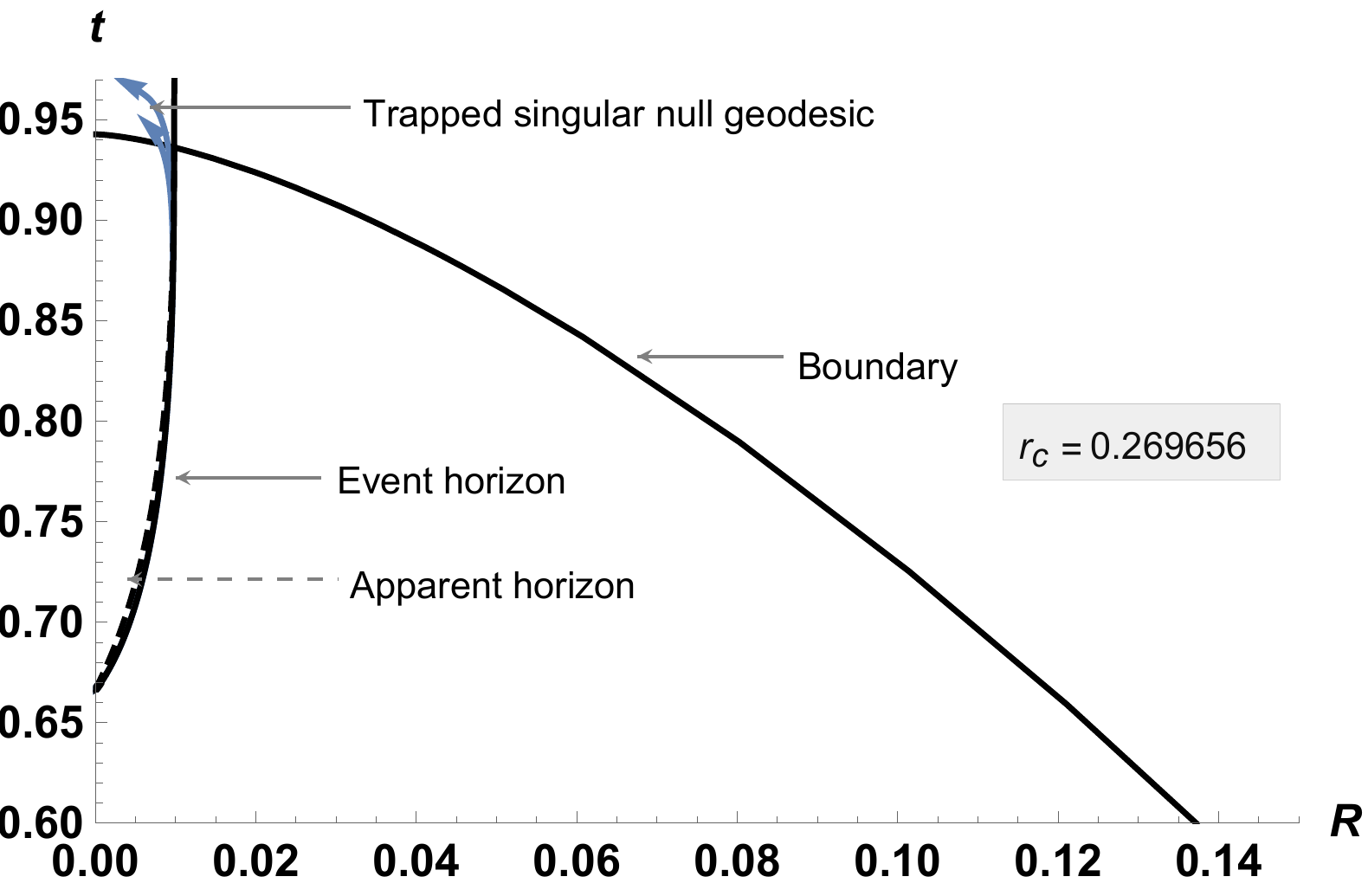}}
\hspace{0.2cm}
\subfigure[]
{\includegraphics[scale=0.54]{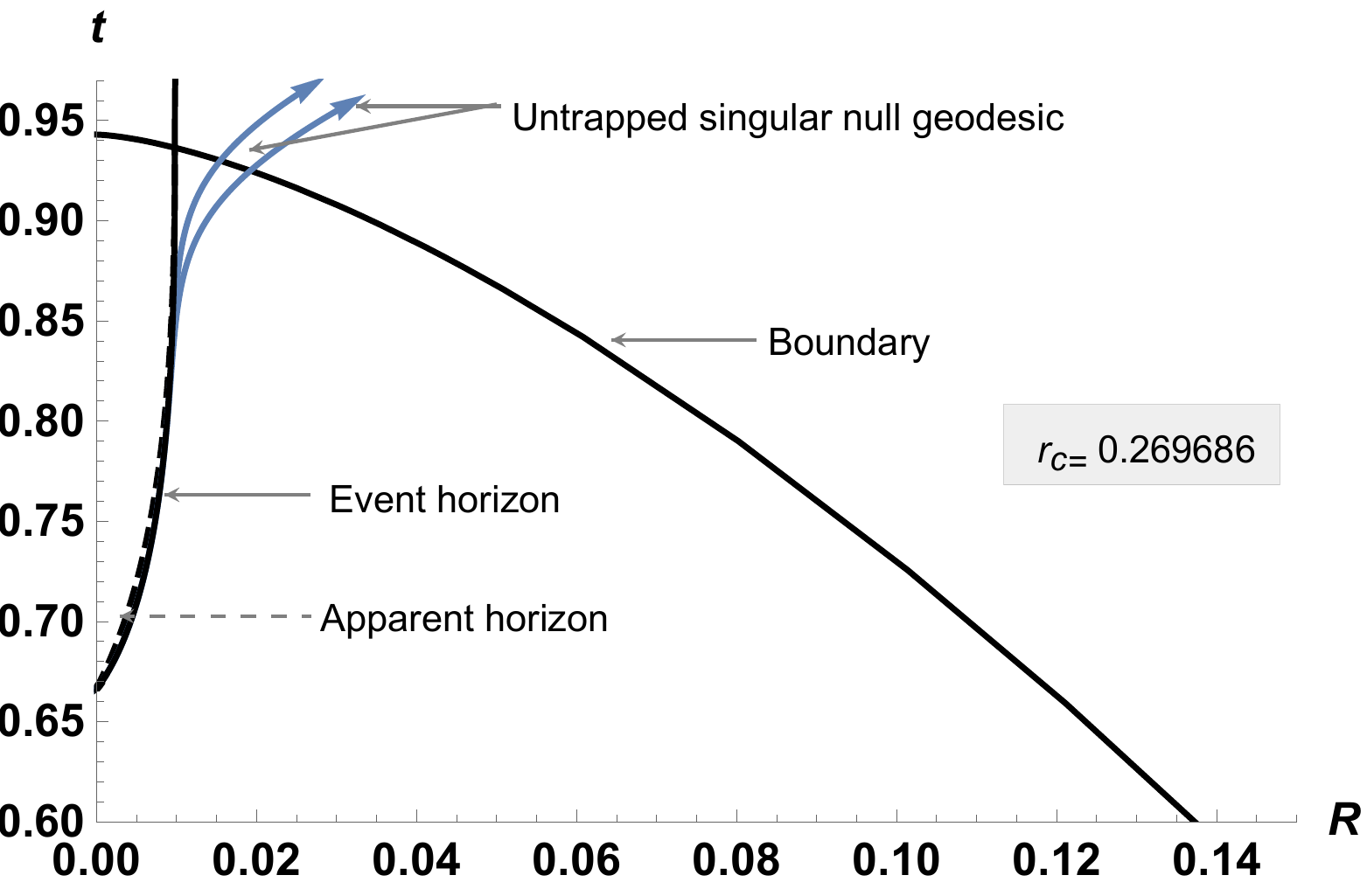}}
\caption{The space-time plot depicting the causal structure of the singularity formed due to marginally bound ($b=0$) collapsing spherical matter cloud with the initial data $S=r^3-25.5r^6$. (a): In GR, the density Eq.(\ref{1efe}) vanishes at $r_c= 0.2696559088937193$. The event horizon forms before the formation of the first singularity, henece the singularity is only locally visible. (b): In $f(R)=R+\alpha R^2$ gravity theory, for $\alpha=10^{-6}$, the density Eq.(\ref{rhofr}) vanishes at $r_c=0.26968557639843954$. The event horizon forms together with the formation of the first singularity, hence the singularity is globally visible.}
\end{figure*}

\begin{figure*}\label{fig4}
{\includegraphics[scale=0.39]{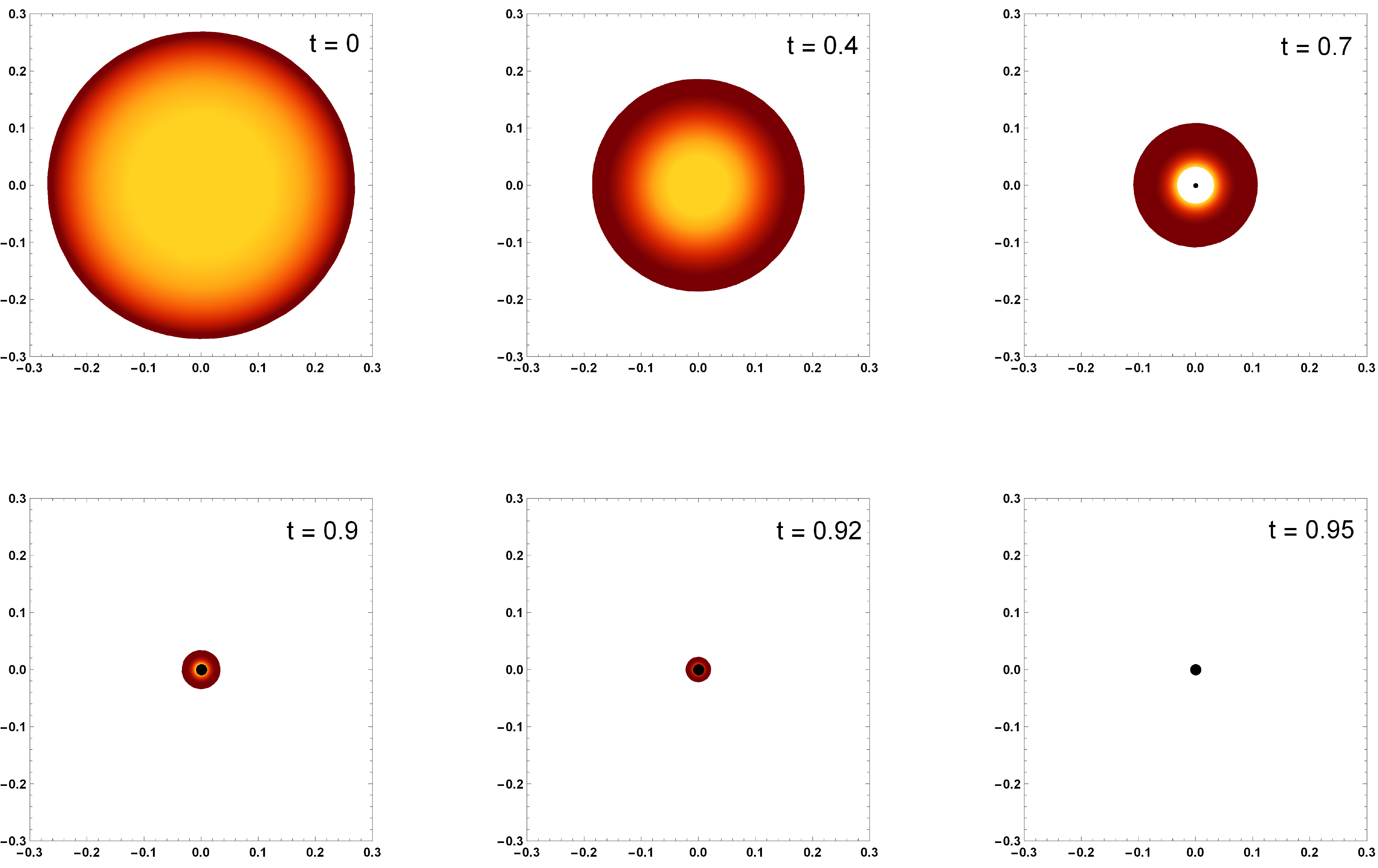}}
\caption{the evolution of the density of the marginally bound collapsing cloud governed by LTB metric and made up of dust with the density Eq.(\ref{1efe}) and the global causal structure of the first central singularity is depicted here. $S=r^3-25.5r^6$. The solid black disk represents the event horizon which increases in size with time and then achieves a fixed physical radius $S(r_c)$. The null geodesic wavefronts  are trapped by trapped surfaces, hence unable to escape from the singular region.}
\end{figure*}
\begin{figure*}\label{fig5}
{\includegraphics[scale=0.39]{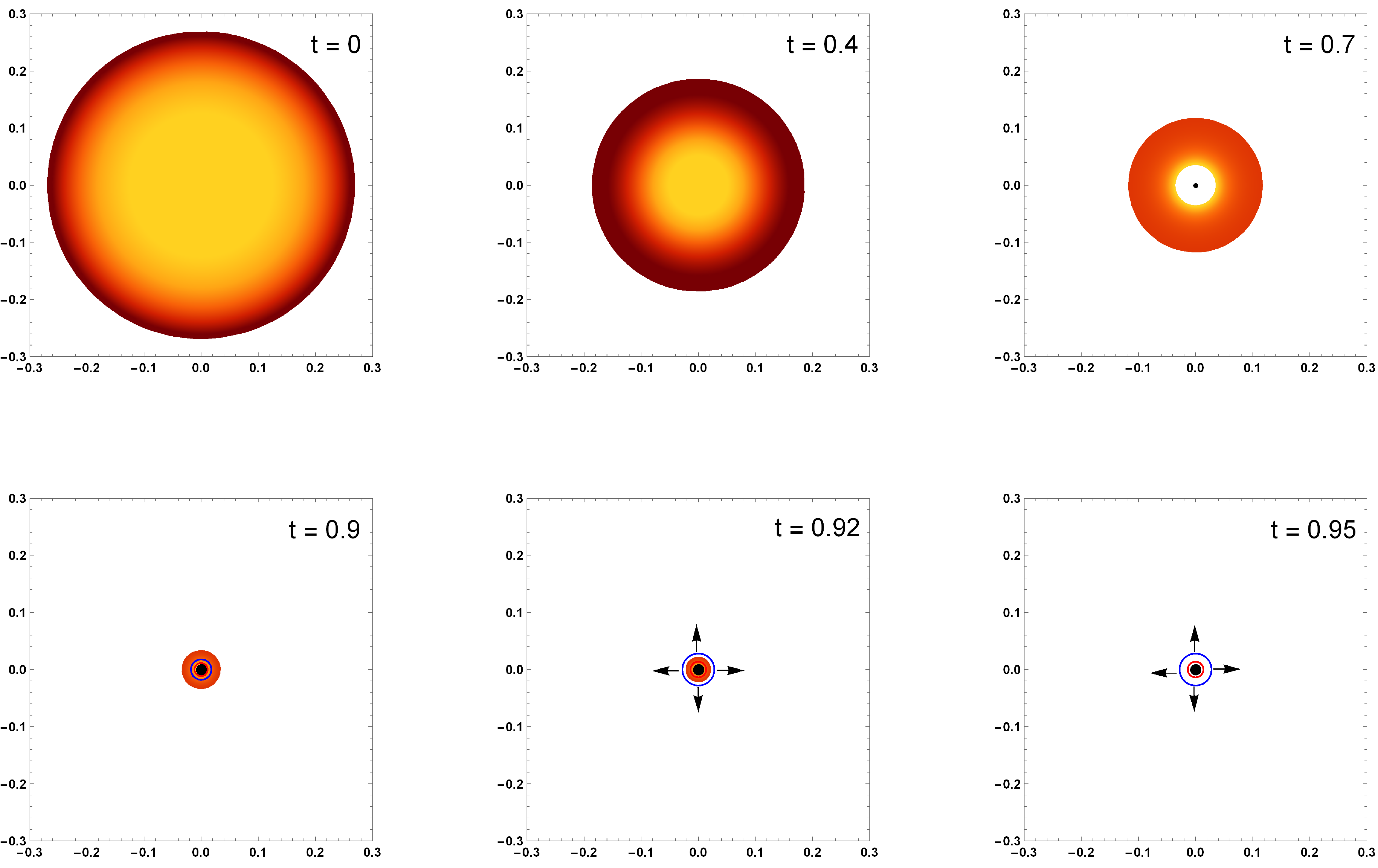}}
\caption{The evolution of the density of the marginally bound collapsing cloud governed by LTB metric and made up of matter field with the profiles expressed in Eq.(\ref{rhofr}-\ref{qfr})  (which is in the framework of $f(R)=R+\alpha R^2$ gravity, with $\alpha=10^{-6}$) and the global causal structure of the first central singularity is depicted here. $S=r^3-25.5r^6$. The solid black disk represents the event horizon which increases in size with time and then achieves a fixed physical radius $S(r_c)$. The null geodesic wavefronts escaping from the singular region are depicted by red and blue concentric circles, which increase in size with time.}
\end{figure*}
The energy density, radial and tangential pressures, and the heat flux of the marginally bound collapsing cloud governed by the LTB metric, Eq.(\ref{LTB}) with $b=0$, are respectively as follows:
\begin{widetext}
\begin{equation}\label{rhofr}
\rho=\frac{S'}{A^2A'} +2\alpha \left(\frac{R \dot A^2}{A^2}+\frac{2R \dot A \dot A'}{A A'}-\frac{R^2}{4}+\frac{2\dot R \dot A}{A}+\frac{\dot R \dot A'}{A'}  -\frac{2R'}{A A'}+\frac{R' A''}{A'^{3}}=\frac{R''}{A'^{2}}\right),
\end{equation}
\begin{equation}\label{prfr}
    p_r=-\frac{\dot S}{A^2 \dot A}+2\alpha\left(-\frac{R\dot A^2 }{A^2}-\frac{2R\ddot A}{A}+\frac{R^2}{4}-\frac{2\dot R \dot A}{A}-\ddot R+\frac{R'}{A A'}\right),
\end{equation}
\begin{equation}\label{ptfr}
    p_t=-\frac{\ddot A}{A}-\frac{\dot A \dot A'}{A A'}-\frac{\ddot A'}{A'}+2\alpha\left(-\frac{R \ddot A}{A}-\frac{R \ddot A'}{A'}-\frac{R \dot A \dot A'}{A A'}+\frac{R^2}{4}-\frac{\dot R \dot A}{A}-\frac{\dot R \dot A'}{A'}-\ddot R+\frac{R'}{A A'}-\frac{R' A''}{A'^{3}}+\frac{R''}{A'^{2}}\right),
\end{equation}
\end{widetext}
\begin{equation}\label{qfr}
    q=\frac{2\alpha}{A'}\left(\dot R'-\frac{R' \dot A'}{A'}\right).
\end{equation}

The function $S$ in the Eq.(\ref{rhofr}-\ref{prfr}), in terms of metric components is expressed as in Eq.(\ref{S}) with $b=0$. However, it is necessary to note that unlike in GR, $S$ no more physically signifies the mass inside a collapsing shell of radial coordinate $r$ at time $t$. It can now be considered just an arbitrary function of the metric components. 

The collapsing matter field should satisfy all the energy conditions throughout the collapse, for which the following inequalities should be satisfied
\cite{Kolassis}:
\begin{equation}\label{ec1}
\vert \rho+p_r \vert-2\vert q \vert \geq 0,
\end{equation}

\begin{equation}\label{ec2}
    \rho-p_r +2p_t+\Delta \geq 0,
\end{equation}

\begin{equation}\label{ec3}
    \rho-p_r +\Delta \geq 0,
\end{equation}
\begin{equation}\label{ec4}
    \rho-p_r \geq 0,
\end{equation}
\begin{equation}\label{ec5}
    \rho-p_r-2p_t +\Delta \geq 0,
\end{equation}
\begin{equation}\label{ec6}
    2p_t +\Delta \geq 0,
\end{equation}
where $\Delta=\sqrt{\left(\rho+p_r \right)^{2}-4 q^2}$.
In Fig.(I), we have depicted the satisfaction of all the above inequalities, for the matter field having density, pressures, and heat flux as in Eq.(\ref{rhofr}- \ref{qfr}), and for certain fixed parametric values, as mentioned in the caption. It should be noted that the energy conditions are also satisfied for all nearby parametric values. Hence, there exists a non-zero measured set of parameters in the density profile and a non-zero range of $\alpha$ such that the energy conditions will not be violated for a small perturbation in these parametric values.


\subsection{Matching condition}

\textcolor{black}{ 
We have a collapsing spherically symmetric marginally bound Lemaitre-Tolman-Bondi cloud that we call here the interior spacetime $\mathcal{V}^{-}\subset \mathcal{M}$, with the metric tensor denoted by $g^-$. In order to study the entire spacetime as a whole (and not just the region $\mathcal{V}^{-}$), we have to take into consideration the spacetime region $\mathcal{M}-\mathcal{V}^{-}$ that we call $\mathcal{V}^{+}$ having the metric tensor that we denote by $g^{+}$. Now, the question that arises is as follows: Can the union of $g^-$ and $g^+$ form a valid solution of the field equations corresponding to $f(R)$ gravity? In other words, what conditions should be imposed on the freedoms available within $g^-$ and $g^+$ so that the regions $\mathcal{V}^-$ and $\mathcal{V}^+$ are joined smoothly at what we call the matching surface $\Sigma\subset \mathcal{M}$ (a three dimensional submanifold)? In general relativity, the smooth matching is achieved by  equating: 1) the induced metric on $\Sigma$ from both the regions: $\mathcal{V}^-$ and $\mathcal{V}^+$, and 2) the extrinsic curvatures of $\Sigma$ as seen from both $\mathcal{V}^-$ and $\mathcal{V}^+$. The first matching condition keeps the first derivative of the metric (made from the union of $g^-$ and $g^+$) from blowing up at $\Sigma$. The second matching condition keeps the energy-momentum tensor from blowing up at $\Sigma$.} 
\cite{Darmois, Israel, Poisson}. 
It can be shown that in GR, the interior LTB spacetime can be matched with the exterior static, spherically symmetric, asymptotically flat vacuum spacetime, which is the Schwarzschild spacetime. However, in the framework of $f(R)$ gravity, for smooth matching of the spacetimes of these two regions, apart from the above-mentioned two junction conditions, the continuity of the  Ricci scalar and its radial derivative at the boundary is also required. 

\textcolor{black}{To make the discussion on matching two spacetime extensive, let us choose a particular coordinate system. Since the extrinsic curvature of a nowhere null hypersurface gives us a good physical intuition in the Gaussian normal coordinates, we choose to work in such coordinate system. We keep a nowhere null hypersurface ($\Sigma$ in our case) as a base to develop the chart in the neighborhood of (a portion of) $\Sigma$. For each $p\in P (\textrm{open})\subset \Sigma$, $\exists$ $\textbf{n}\in T_p\mathcal{M}$ such that $\textbf{n}$ is orthogonal to all vectors in $T_p\Sigma$, and $g(\textbf{n},\textbf{n})=1$. Now, for each $p\in P$, $\exists$ a unique geodesic with tangent $\textbf{n}$, and affine parameter $l$ such that $\gamma(l=0)=p$.  For a predefined chart $\phi:P\to \mathbb{R}^3$, each point $q$ lying on $\gamma$ has coordinates $(l,\phi(p))$, where $\gamma(l)=q$. This way, we can define a coordinate system in the neighborhood of $P$.  In the Gaussian coordinate basis, we write the interior and the exterior metric as
\begin{equation}
    ds^2=dl^2+ \eta_{ij}d\zeta^id\zeta^j
\end{equation}
where $\zeta^i$ are the coordinates on $\Sigma$, and $\Sigma$ corresponds to $l=0$. The extrinsic curvature of $\Sigma$ in arbitrary coordinates is defined as
\begin{equation}
    K_{ij}= \mathcal{L}_{\textbf{n}}g_{ij},
\end{equation}
i.e. the Lie derivative of the metric tensor with respect to the vector field $\textbf{n}$.
In the Gaussian normal coordinate system, it is written in terms of the usual derivative of the induced metric components (on $\Sigma$) with respect to $l$ as
\begin{equation}
    K_{ij}=\frac{1}{2}\frac{\partial \eta_{ij}}{\partial l},
\end{equation}
which helps us interpret the extrinsic curvature of $\Sigma$ as a measure of change of the induced metric with respect to its normal.  
The Ricci scalar $R$ is then written as
\begin{equation}
    R=2\partial_l K-\Tilde{K}_{ij}\Tilde{K}^{ij}-\frac{4}{3}K^2+\mathcal{R}.
\end{equation}
Here, $K$ is the trace of the extrinsic curvature, and $\Tilde{K}_{ij}$ is its trace-free part, and $\mathcal{R}$ is the Ricci scalar of the hypersurface $\Sigma$, constructed from the components of the three-metric $\eta_{ij}$.}

\textcolor{black}{
The matching conditions in the domain of general relativity is expressed as
\begin{equation}\label{jc1}
    \left[\eta_{ij}\right]^+_-=0,
\end{equation}
\begin{equation}\label{jc2}
    \left[\Tilde{K}_{ij}\right]^+_-=0,
\end{equation}
and
\begin{equation}\label{jc3}
    \left[K\right]^+_-=0.
\end{equation}
\textcolor{black}{Extension of the junction conditions for smooth matching of two disjoint spacetime is carried out in different frameworks beyond GR (see for e.g. 
\cite{Deruelle_2008, Senovilla_2013, Olmo_2021, Rosa_2021, Rosa_2022}).}
In the case of general $f(R)$ gravity, the additional matching conditions are
\begin{equation}\label{jc4}
    \left[R\right]^+_-=0,
\end{equation}
and
\begin{equation}\label{jc5}
    \left[\partial_l R\right]^+_-=0.
\end{equation}
}
It was shown by Nzioki \textit{et. al} 
\cite{Nzioki} 
that for a class of $f(R)$ gravity model, which includes the Starobinsky one, the Schwarzchild solution is the only static, spherically symmetric, asymptotically flat vacuum spacetime with vanishing Ricci scalar, thereby extending the Birkhoff's theorem in $f(R)$ gravity. \textcolor{black}{Hence, we match the interior LTB modeled collapsing cloud $\mathcal{V}^-$ with the exterior Schwarzschild spacetime $\mathcal{V}^+$ given by
\begin{equation}
    ds^2=-\left(1-\frac{2m}{r_s}\right)dT^2+\frac{dR^2}{\left(1-\frac{2m}{r_s}\right)}+r_s^2 d\Omega^2,
\end{equation}
using the conditions (\ref{jc1}-\ref{jc5}). Here $r_s$ is the Schwarzschild radius and $(T,R)$ are the Schwarzschild coordinates. The condition (\ref{jc1}) imply that
\begin{equation}
    r_s=A(t,r).
\end{equation}
The condition (\ref{jc2}) and (\ref{jc3}) imply that on $\Sigma$
\begin{equation}
    G^1_1\vert_{\Sigma}=0
\end{equation}
as seen from $\mathcal{V}^-$. For the LTB spacetime, $G^1_1=0$ throughout $\mathcal{V}^-$, and hence the conditions (\ref{jc2}) and (\ref{jc3}) are naturally satisfied.}

Goswami \textit{et. al.} 
\cite{Goswami} 
in 2014 showed that \textcolor{black}{satisfying the conditions (\ref{jc4}) and (\ref{jc5})}  constrains the previously free function $S(r)$, thereby fine-tuning it and making it unstable under matter perturbation. The argument goes as follows: The Ricci scalar is expressed in terms the arbitrary function $S$ as
\begin{equation}\label{R1}
    R=-\frac{3M+rM'}{ v^2\left(rv'+v \right)},
\end{equation}
where $M$ is related to $S$ as $M=\frac{S}{r^3}$, and $v=v(t,r)=\frac{A}{r}$ is called the scaling function. The scaling function can be thought of as the redefined time coordinate such that at the initiation of the collapse, $v(0,r)=1$ for all the shells, and $v(t,r)=0$ when the shell of comoving radius $r$ collapses to a singularity. \textcolor{black}{Now, in order for the smooth matching of the Ricci scalar and its derivative $\partial_l R$ of the interior LTB cloud with that of the exterior Schwarzschild spacetime at the junction,  both the quantities} for the interior spacetime should vanish at the boundary $r_c$. Hence $R$ should have the form
\begin{equation}\label{R2}
    R=(r_c-r)^{2}g(t,r).
\end{equation}
Equating Eq.(\ref{R1}) and Eq.(\ref{R2}), we obtain
\begin{equation}\label{odem}
    rM'+3M=j(r) (r_c-r)^2,
\end{equation}
where
\begin{equation}
    j(r)=v^2(rv'+v)g(t,r).
\end{equation}
The functional form of $v$ and $g(t,r)$ is determined once the functional form of $M$ is determined. Hence we can say that $j=j(M,r)$. The Eq.(\ref{odem}) then becomes a first-order ordinary linear differential equation, which can only be satisfied by a class of functions $M$. This is how the additional matching condition constrains the function $S$ in \textcolor{black}{$f(R)$} theories of gravity, which was free to choose in GR. 

\textcolor{black}{Senovilla
\cite{Senovilla_2013}
in 2013 showed that in the special case where $d^3f(R)/dR^3=0$, a discontinuity in the Ricci scalar is permitted (this discontinuity comes with a sound physical interpretation of energy-momentum tensor on $\Sigma$). Hence, as far as $f(R)=R+\alpha R^2$ gravity is concerned, the only conditions that should be of concern are Eq.(\ref{jc1}-\ref{jc3}), and one should not bother about conditions (\ref{jc4}) and (\ref{jc5}).} 

\subsection{Globally visible nodal singularity in f(R) gravity.}

The collapse formalism of the LTB metric in GR discussed in the previous section, i.e. Eq.(\ref{S}) and Eq.(\ref{t-ts}-\ref{teh}), is same for the collapsing matter field governed by identical LTB metric, in $f(R)$ gravity for a time-independent function $S$ and vanishing $b$. However, the boundary $r_c$ of the collapsing cloud, which we define as the comoving radius where $\rho$ vanishes, will be different in $f(R)$ gravity. This is because the density profile Eq.(\ref{rhofr}) is different from Eq.(\ref{1efe}). This causes a change in the evolution of the event horizon, thereby affecting the global causal structure of the singularity. The difference can be clearly seen in Fig.(II). In this figure, the geometry governing the collapse of two different matter fields in two different theories of gravity is the same. By this, we mean that apart from both matter fields being governed by the LTB metric, the initial data $(S,b)$ is also the same in both cases. However, since the boundaries of these two collapsing clouds are different, the initial condition Eq.(\ref{teh}), which needs to be satisfied by the solution of the differential Eq.(\ref{deteh}) for it to represent the dynamics of the event horizon, is changed. Hence the previously locally visible singularity in GR is now globally visible in $f(R)$ gravity. The evolution of the density of the matter field along with the trapped (in GR) and escaped (in $f(R)$) null geodesics are depicted in Fig.(III) and Fig.(IV), respectively. For a fixed functional form of $S$ as mentioned in the captions, for $\alpha=10^{-6}$, one gets a globally visible singularity. However, this is not the only value. One can show that for any greater value of  $\alpha$, globally visible singularity is achieved. This means that one can trace infinite event horizons, each corresponding to one value of $\alpha$, which are solutions of the differential equation (\ref{deteh}) and starting from $(t_s(0),0)$ in the $(t,r)$ plane. This is only possible if $(t_s(0),0)$ is a nodal point.

In order to check if the first singularity is a nodal point, consider two different frameworks of gravity, both of which are Starobinsky type, but with different values of scalar multiples non-minimally coupled with the quadratic curvature term in the Lagrangian. Let us call them $\alpha_1$ and $\alpha_2$, with
\begin{equation}
\alpha_2>\alpha_1     
\end{equation}
($\alpha_1$ can also be zero, which corresponds to GR). The evolution of the two distinct event horizons, each corresponding to distinct values of $\alpha$, are dictated by differential Eq.(\ref{deteh})
respectively satisfying 
\begin{equation}
    t_{EH}(r_{1})= \frac{2}{3}\frac{r_{1}^{\frac{3}{2}}}{\sqrt{S(r_1)}}-\frac{2}{3}S(r_1), 
\end{equation}
and
\begin{equation}
    t_{EH}(r_{2})= \frac{2}{3}\frac{r_{2}^{\frac{3}{2}}}{\sqrt{S(r_2)}}-\frac{2}{3}S(r_2).
\end{equation}
Here $r_1$ and $r_2$ are the largest comoving radius of the collapsing cloud corresponding to $\alpha_1$ and $\alpha_2$ respectively. Let us choose $\alpha_1$ such that for a given fixed functional form of $S$, the first singularity is globally visible. One can therefore see that for $\alpha=\alpha_1$ at $(t_{s}(0),0)$ in the $(t,r)$ plane, 
\begin{equation*}
    \frac{dt_{EH}(r)}{dr}
\end{equation*}
is not continuous. This is because for a small change in $t_{EH}$ from $t_{s}(0)$ to some $t_f$ where 
\begin{equation}
    t_f>t_s(0),
\end{equation}
there is zero change in $r$, since all null geodesics at $r=0$ are trapped after the time $t_s(0)$. Hence,  $\frac{dt_{EH}}{dr}$ is infinite at $r=0$.

Now, the uniqueness theorem of the first order linear differential equation says that if 
\begin{equation}
    g(x,y)\hspace{1cm} \textrm{and} \hspace{1cm} \frac{\partial g(x,y)}{\partial y}
\end{equation}
are continuous in the neighborhood around $x=0$, then the solution (in a possibly smaller neighborhood around $x=0$) of the differential equation (with initial condition) given by
\begin{equation}
    y'=g(x,y), \hspace{1cm} y(x_0)=y_0 
\end{equation}
is unique. However, this uniqueness theorem is not applicable in our case because of the discontinuity of $A'(t,r) (=\frac{dt_{EH}}{dr})$ at $(t_s(0),0)$. Therefore, one can have more than one solution of the differential Eq.(\ref{deteh}), and passing through $(t_s(0),0)$, making it a nodal point. 

One can check numerically that the singularity is globally visible for any $\alpha=\alpha_2>\alpha_1$, if it is globally visible for $\alpha=\alpha_1$. This supports the claim that the first central singularity is indeed a nodal point.

Let us now fix the framework of gravity. For the singularity to be visible by an asymptotic observer for infinite time, the central singularity should emit a congruence of infinite null geodesics, each redshifted by a different amount and the event horizon being the most redshifted (infinitely) null geodesic. This can happen because we have shown that $(t_s(0),0)$ is a nodal point.

\section{Conclusions}
The concluding remarks are as follows:
\begin{enumerate}
\item In order to determine the global causal property of the singularity formed due to a collapsing spherically symmetric matter cloud, only knowing the spacetime metric governing the collapsing matter field is not sufficient. One also has to have the information of the extent to which the spacetime is governed by a given metric. In other words, one also has to have the information of the largest comoving radius $r_c$, which is also the initial size of the collapsing cloud. This boundary of the cloud affects the evolution of the event horizon in that it provides an initial condition to the differential equation whose solution satisfying this initial condition represents the dynamics of the event horizon. 

\item To show this, we considered the same spacetime metric (marginally bound LTB using up the remaining one-degree freedom by fixing the functional form of $S$) governing two different matter fields respectively collapsing unhindered in two different theories of gravity. The LTB metric corresponding to dust in GR corresponds to imperfect viscous fluid in  $f(R)=R+\alpha R^2$ gravity. Since the density profiles of the two clouds are different, their boundaries (which are determined by vanishing density) are also different. For this fixed metric with no remaining functional freedom of choice,  the event horizon, therefore, forms before the formation of the first singularity in GR, but forms together with the formation of the first singularity in $f(R)$ gravity, thereby making the singularity locally visible in GR but globally visible in $f(R)$ gravity. 

\item It should be noted that when we say: ``same spacetime metric in two different theories of gravity" or ``same geometry in two different theories of gravity," we don't mean that the two spacetimes are isomorphic to each other. They are not isomorphic because of the difference in the matching surfaces in both cases. This difference is because we have defined the boundary of the collapsing cloud such that it has the physical radius corresponding to that comoving radius where the density vanishes. These comoving radii are different because the density profiles of the matter fields are different for different theories of gravity.

In scenarios where the collapsing cloud is such that its density does not vanish but has some known value $\rho_c$ at the boundary, the outermost comoving radius $r_c$ in $f(R)$ gravity will still be different from that in GR for the LTB cloud. This difference in $r_c$ causes the event horizons to evolve differently in different gravity theories, possibly affecting the global causal structure of the first central singularity. This is similar to the case of vanishing density at the boundary, which we have considered.  

\item The local causal structure of the singularity is, however, only determined by the behavior of the apparent horizon, which is the boundary of all trapped surfaces, and whose dynamics are completely determined once the governing spacetime metric is known.  

\item Matching the interior collapsing spacetime with the exterior spacetime in the framework of $f(R)$ theories of gravity impose a restriction on the otherwise free function $S$. The spacetime is singular at the matching surface if the junction conditions are violated. We have, however, chosen a specific form of the function $S$, which is $S=r^3-25.5r^6$, as an example to show the difference in the global causal structure of the singularity in GR and in $f(R)$ gravity. Whether or not this specific functional form maintains the continuity of the Ricci scalar and its radial derivative has not been investigated. However, even if there is a jump in the curvature term at the boundary, one could physically interpret this violation of the junction condition such that there exists surface stress-energy term on the matching hypersurface and should not be considered unphysical. 

\item For an asymptotic observer to be able to observe the singularity, apart from the event horizon to form with the formation of the first singularity, the singularity should also be a nodal point. Here we have argued that because of the discontinuity of the function $A'(t,r)$ at $(t_s(0),0)$, the uniqueness theorem of the first-order differential equation does not hold. Hence, there can exist more than one solutions of null geodesic equation starting from $(t_s(0),0)$. We have argued that there indeed exists more than one outgoing singular null geodesics, using the apparent property that the global causal structure of the singularity is stable under small perturbation in the value of $\alpha$, which can be verified numerically.

\item It should be noted that studying the global causal structure of the first central singularity formed due to a collapsing cloud requires the explicit expression of the physical radius in terms of $t$ and $r$. This can be easily obtained in GR in the case of dust collapse using Eq.(\ref{t-ts}) and Eq.(\ref{ts}) to obtain
\begin{equation}
    A(t,r)=\left(r^{\frac{3}{2}}-\frac{3}{2}\sqrt{S}t\right)^{\frac{2}{3}}.
\end{equation}
However, in GR, in the case of the cloud having non-zero pressure, such explicit expression of $A(t,r)$ is difficult to obtain since integrating the analogous equation of Eq.(\ref{S}) is not so straightforward. One way to interpret the resulting global visibility, which we show in $f(R)$ is that the LTB metric in $f(R)$ theory governs a collapsing cloud having some pressure as seen in  Eq.(\ref{prfr}) and Eq.(\ref{ptfr}). This global visibility seems to be generic in nature as far as small perturbations in the initial data $(S,b)$ are concerned. Now, it seems fairly reasonable to assume that even in GR, one should get a non-zero measured set of initial data for which the end state of a ``pressured" collapsing cloud ends up in a globally visible singularity.

\end{enumerate}
\section{Acknowledgement}

KM would like to acknowledge the support of
the Council of Scientific and Industrial Research (CSIR, India, Ref: 09/919(0031)/2017-EMR-1) for funding the work. We thank the reviewer for suggestions, thereby making the article clearer.

\end{document}